\documentclass[preprint,aps,amsmath,superscriptaddress,nofootinbib,tightenlines,showpacs]{revtex4}
\usepackage{amsmath}
\usepackage{amsfonts}
\usepackage{amssymb}
\usepackage{latexsym}
\usepackage{graphicx}
\usepackage{bm}
\usepackage{epsfig}
\usepackage[english]{babel}

\def\horava{Ho\v{r}ava}

\def\mb{\mathbf}
\def\nn{\nonumber}
\def\hl{Ho\v{r}ava-Lifshitz}
\def\p{\partial}
\def\dij{\delta_{ij}}

\begin{document}


\title{Scale Invariant Power Spectrum in Ho\v{r}ava-Lifshitz Cosmology without Matter}
\author{Bin Chen\footnote{Electronic address: bchen01@pku.edu.cn},\hspace{2ex}Shi Pi\footnote{Electronic address: spi@pku.edu.cn}
and Jin-Zhang Tang\footnote{Electronic address: alberttjz@163.com}}
\affiliation{Department of Physics, and State Key Laboratory of
Nuclear Physics and Technology, Peking University, Beijing 100871,
China}

\date{\today\\ \vspace{1cm}}
\begin{abstract}
In this paper we investigate the physical implications of the
dynamical scalar mode in pure \horava-Lifshitz gravity on
cosmology. We find that it can produce
 a scale-invariant power spectrum in UV era if the detailed balance condition on the action is
 relaxed. This indicates that the physical scalar mode may seed the large scale
 structure and \hl~cosmology could be a qualified alternative to
 inflationary scenario.


\end{abstract}
\maketitle

\section{introduction}\label{sec-intro}

Recently, inspired by the quantum critical phenomena in condensed
matter physics\cite{Quantumphase}, \horava~proposed a
renormalizable gravitation theory with dynamical critical exponent
$z=3$ in UV limit\cite{Horava:2008ih,Horava:2009uw}. In this
\horava-Lifshitz theory, the time and space will take different
scaling behavior as
\begin{equation}\label{scaling}
    \mb{x}\rightarrow b\mb{x},\;\;\;\; t\rightarrow b^zt,
\end{equation}
where $z$ is the dynamical critical exponent characterizing the
anisotropy between space and time.  In the action, the kinetic
term is still quadratic in the first derivatives of the metric of
spatial slice, but the potential terms may have higher order
spatial derivatives. This fact help us avoid the breaking of
unitarity, which happens in other higher derivative gravity
theory. Due to the existence of higher spatial derivative terms,
the UV behavior of the theory is much improved. In fact, from
power counting, it seems that the theory is renormalizable in the
UV. By turning on relevant perturbation, the theory flows to an IR
fix point where it behaves like standard Einstein general
relativity. As a price one has to pay, the Lorentz symmetry is
broken by construction and it will only emerge as an accidental
symmetry in the IR.

Since its discovery, many authors have attempted to apply this
\hl~gravity to the study of early
cosmology\cite{Calcagni:2009ar,Kiritsis:2009sh}. Especially it was
pointed out that the \hl~gravity may be a candidate for theory of
the early universe instead of inflation. The casuality and
flatness problem could be solved under this framework as in
inflationary model.
By considering the scalar field coupled to gravity, it was shown
in
 \cite{Calcagni:2009ar,Kiritsis:2009sh,Mukohyama:2009gg},
that  the scale-invariant power spectrum can be produced, with a
new dispersion relation at UV
\begin{equation}\label{dispersion:Lifshitz scalar}
E\sim k^6.
\end{equation}
The key point here is that the matter scalar action may have
higher derivative terms, allowed by the scaling (\ref{scaling}).
Such scalar field theory could be taken as field theory at
Lifshitz points without detailed balance condition\cite{Lifshitz}.
The general field theory at a Lifshitz point was discussed in
\cite{Chen:2009ka}. The other applications of \hl~gravity to
cosmology and black hole physics could be found in \cite{HL}.

 On the other hand, one important feature of \hl~gravity is that due to anisotropy,
 the usual diffeomorphism is broken to foliation-preserving diffeomorphisms. As a result,
 there exist an extra dynamical scalar degree of freedom. This fact has been shown both for the perturbations
 about flat background\cite{Horava:2009uw} and about FRW cosmology\cite{Cai:2009dx}.
   This means that we can have a scalar mode from the
perturbation of gravitational field in the early universe, besides
the other two tensor modes of gravitational waves. It would be
interesting to investigate the physical implication of such scalar
mode. 

In this paper, we consider the possibility if we may interpret the
scalar mode of the gravity as the origin of the large scale
structure and cosmic microwave background anisotropy in the IR
era, without any additional matter field. If this is case, we need
neither the inflaton nor curvaton in the early universe, and
gravity itself can produce the seed of the structure. As the first
step, we need to check if this scalar mode could produce the
scale-invariant spectrum. But unfortunately, in \cite{Cai:2009dx},
the dispersion relation of the gravitational scalar is no more
\eqref{dispersion:Lifshitz scalar}, but instead
\begin{equation}\label{dispersion:Cai}
    E^2\sim k^4.
\end{equation}
Thus the power spectrum will be proportional to $k$ which is not
consistent with  the CMB observation. The essential point is that
the relation \eqref{dispersion:Cai} stems from the fact that the
terms of sixth order derivatives in the equation of motion are
absent due to the detailed balance condition, which was used
originally by \horava~ to simplify the form of the action of
gravity. We will show that after we relax the detailed balance
condition,  we can recover the scale invariant spectrum for the
gravitational scalar.

The rest of this paper is organized as follows. In section
\ref{sec-review} we will take a brief review of the \hl~gravity,
while in section \ref{sec-calcu} we study the cosmological
implication of the gravitational scalar. After relaxing the
detailed balance condition and doing analytic continuation of two
parameters, we obtain the action of the gravity, which are
slightly different from the original action but allows us to have
sixth order spatial derivative in the equation of motion of scalar
mode. We derive the equation of motion of the gravitational scalar
without detailed balance condition and calculate the scalar power
spectrum. In section \ref{sec-inclu} we will end with some
discussions.

\section{Review of \horava-Lifshitz gravity with detailed balance
condition}\label{sec-review}

In this section, we give a brief review of the \horava-Lifshitz
gravity.  Due to the anisotropy of time and space, it is more
convenient to work  with the ADM metric,
\begin{equation}\label{ADMmetric}
    ds^2=-N^2dt^2+g_{ij}(dx^i+N^idt)(dx^j+N^jdt).
\end{equation}
The classical scaling dimensions of the fields are
 \begin{equation}
 [N]=0, ~~~[N_i]=z-1,~~~[g_{ij}]=0.
 \end{equation}  From this
metric, we can construct an action of gravity from the scaling
rule \eqref{scaling} and the invariance under foliation-preserving
diffeomorphisms. It turns out that the kinetic term is quadratic
in the second fundamental form. And before considering the
relevant terms of lower scaling dimension, we consider the
marginal potential term which dominate on the high-energy limit.
In $z=3,D=3$ case,  it could contain the higher derivatives of the
metric up to sixth order\cite{Horava:2009uw}. The terms quadratic
in $g_{ij}$ could be
\begin{equation}\label{type of R's}
    \nabla_iR_{jk}\nabla^iR^{jk},\;\;\;\;\nabla_iR_{jk}\nabla^kR^{ij},\;\;\;\;R\nabla^2R,\;\;\;\;R_{ij}\nabla^2R^{ij}.
\end{equation}
The other combinations are either identical up to a total
derivative or related by Bianchi identity or other symmetries.
Although there are many constraints, the relative magnitudes of
these terms are arbitrary. The other marginal terms are cubic in
curvature. For the relevant terms, there are even more
possibilities. To reduce these uncertainties \horava~suggest to
impose the detailed balance condition to confine the action
further more. Under this condition, the potential term including
relevant perturbations are determined by
\begin{equation}
S_v=\frac{\kappa^2}{8}\int dtd^3x\sqrt{g}E^{ij}{\cal
G}_{ijkl}E^{kl},
\end{equation}
where
\begin{equation}
\sqrt{g}E^{ij}=\frac{\delta W[g_{kl}]}{\delta g_{ij}}
\end{equation}
for some action $W$ in three-dimension and ${\cal G}_{ijkl}$ is
the De Witt metric defined by
\begin{equation}
{\cal G}^{ijkl}=\frac{1}{2}(g^{ik}g^{jl}+g^{il}g^{jk})-\lambda
g^{ij}g^{kl}.
\end{equation}
The action $W$ could be of the form
\begin{equation}
W=\frac{1}{\omega^2}\int \omega_3(\Gamma)+\mu \int d^3x
\sqrt{g}(R-2\Lambda_W),
\end{equation}
where $\omega_3$ is the gravitational Chern-Simons term and
$\omega$ is a dimensionless coupling constant. $\mu$ is coupling
constant with dimension $[\mu]=1$ and $\Lambda_W$ is effectively a
cosmological constant with dimension $[\Lambda]=2$.

Finally the action of gravity with detail balance can be written
as
\begin{eqnarray}\label{S_g}\nn
    S_g&=&\int dtd^3\mb{x}\sqrt{g}N\left\{\frac{2}{\kappa^2}(K_{ij}K^{ij}-\lambda K^2)
    -\frac{\kappa^2}{2\omega^4}C_{ij}C^{ij}+\frac{\kappa^2\mu}{2\omega^2}\frac{\epsilon^{ijk}}{\sqrt{g}}
    R_{il}\nabla_jR^l{}_k\right.\\
    &&\left.-\frac{\kappa^2\mu^2}{8}R_{ij}R^{ij}
    +\frac{\kappa^2\mu^2}{8(1-3\lambda)}\left(\frac{1-4\lambda}{4}R^2
    +\Lambda_WR-3\Lambda_W^2\right)\right\},
\end{eqnarray}
where $K_{ij}$ is the second fundamental form, or the extrinsic
curvature, of the spatial slice; $C_{ij}$ is the Cotton tensor
which is used to construct the action preserving the detailed
balance condition; and $R_{ij}$ is the Ricci tensor in the three
dimensional space. Their definitions are
\begin{eqnarray}\label{K}
K_{ij}&=&\frac{1}{2N}(\dot{g}_{ij}-\nabla_iN_j-\nabla_jN_i),\\\label{C}
C_{ij}&=&\frac{\epsilon^{ikl}}{\sqrt{g}}\nabla_k\left(R^j{}_l-\frac{1}{4}R\delta^j_l\right).
\end{eqnarray}
The first parenthesis in \eqref{S_g}  involving only the extrinsic
curvature is the kinetic term, while the others are potential
terms. $\lambda$ is the coupling constant in kinetic term, and
runs to $\lambda=1$ in IR era such that the kinetic term goes back
to the case in general relativity.  $\kappa$ is something like the
Newton's gravitational constant $G_N$, and indeed there is a
proportional relation between them
 \begin{equation}
 G_N=\frac{\kappa^2}{32\pi c}
 \end{equation}
 where the speed of light is
 \begin{equation}
 c=\frac{\kappa^2\mu}{4}\sqrt{\frac{\Lambda_W}{1-3\lambda}}.
 \end{equation}

Note that in UV era, the dominant potential term will be the
integral of
\begin{equation}\label{Svuv,with detailed balance}
C_{ij}C^{ij}
=(\nabla_iR_{jk}\nabla^iR^{jk}-\nabla_iR_{jk}\nabla^jR_{ik}-\frac{1}{8}\nabla_iR\nabla^iR),
\end{equation}
which is the only term involving sixth order derivatives of the
metric after partial integrations. The Cotton tensor $C_{ij}$ is
symmetric and traceless, and is covariantly conserved. Most of
all, it is conformal,
\begin{equation}\label{conformalCotton}
    g_{ij}\rightarrow\exp(2\Omega(\mb{x}))g_{ij},\;\;\;\;C_{ij}\rightarrow\exp(-5\Omega(\mb{x}))C_{ij}.
\end{equation}
And it is the successor of Weyl tensor in three dimensional space
to play the role of the criterion of conformal flatness.


\section{the power spectrum without detail balance}\label{sec-calcu}


In this section we will calculate the equation of motion of the
extra scalar mode in \hl~gravity in the early universe without
matter, but the detailed balance condition will be relaxed. We
begin with perturbed Friedmann-Robertson-Walker metric in ADM form
after gauge fixing\cite{Cai:2009dx}
\begin{equation}\label{metric}
    ds^2=-c^2dt^2+a^2(1-2\psi)\delta_{ij}(dx^i+a^{-1}\p^iBdt)(dx^j+a^{-1}\p^jBdt),
\end{equation}
where $\delta_{ij}$ is the metrc on the three dimentional leaf of
the foliation. Here we have assumed the background universe is
flat. From (\ref{metric}),  we have
\begin{eqnarray}
 N^2&=&c^2, \\
  N_i&=&ca\p_iB \\
  g_{ij}&=&a^2(1-2\psi)\delta_{ij}\simeq a^2e^{-2\psi}\dij,\\
  g^{ij}&\simeq& a^{-2}(1+2\psi)\delta^{ij}\simeq a^{-2}e^{2\psi}\delta^{ij},
\end{eqnarray}
where $\simeq$ means equalization accurate up to first order, and we
adopt the exponential expression of the scalar $\psi$ for
convenience, as in \cite{Maldacena:2002vr}. From this metric we have
the Ricci tensor and scalar,
\begin{eqnarray}\label{Rij}
  R_{ij} &=& [\p^2\psi-(\p\psi)^2]\dij+\p_i\p_j\psi+\p_i\psi\p_j\psi, \\\label{R}
  R &=& 2a^{-2}e^{2\psi}[2\p^2\psi-(\p\psi)^2].
\end{eqnarray}
More importantly the spatial slice is conformal flat. This fact
leads to vanishing Cotton tensor and the absence of six derivative
terms in the equation of motion of $\psi$. The same thing happens
for the perturbation around the flat spacetime background. In the
latter case, the equation of motion of physical scalar mode
contains terms with spatial derivatives up to fourth order. As a
result, the ultraviolet behavior is not good enough. This stems
from the detailed balance condition.

As pointed out in the original paper \cite{Horava:2009uw}, there
is no first principle at this moment to decide which kind of
combination of terms in (\ref{type of R's}) is more physical. The
coefficients of these terms could be arbitrary, only being
constrained by the requirements of stability and unitarity of the
quantum theory. Just from simplicity, the detailed balance
condition was imposed. The very interesting relation to
three-dimensional massive gravity from detailed balance is another
story, being very little  to do with the renormalizability of the
theory. Therefore, one can expect that without detailed balance
condition the theory is still UV well-defined. In the following we
will take this philosophy and consider the more general action
without detailed balance.

Our starting point is the following action of the pure \hl~~
gravity
\begin{equation}\label{S=Sk+Svuv+Svinter+Svir}
    S=S_K+S_V^{\text{(UV)}}+S_V^{(\text{inter})}+S_V^{(\text{IR})},
\end{equation}
where $S_K$ is the kinetic term, $S_V^{\text{(UV)}}$,
$S_V^{(\text{inter})}$ and $S_V^{(\text{IR})}$ are terms dominant
in UV, intermediate and IR era, respectively, being of the forms
\begin{eqnarray}\label{Sk}
S_K &=& \int dtd^3\mb{x} \sqrt{g}N\alpha(K_{ij}K^{ij}-\lambda
K^2),\\\label{Svuv} S_V^{\text{(UV)}}&=&\int{}dtd^3\mb{x}
\sqrt{g}N\beta(\beta_1\nabla_iR_{jk}\nabla^iR^{jk}+\beta_2\nabla_iR_{jk}\nabla^jR^{ik}+\beta_3\nabla_iR\nabla^iR),
\\\label{Svinter}
S_V^{(\text{inter})}&=&\int{}dtd^3\mb{x}
\sqrt{g}N\left\{\gamma\frac{\epsilon^{ijk}}{\sqrt{g}}R_{il}\nabla_jR^l{}_k+\zeta{}R_{ij}R^{ij}+\eta{}R^2\right\},
\\\label{Svir}
S_V^{(\text{IR})} &=& \int dtd^3\mb{x} \sqrt{g}N(\xi R+\sigma),
\end{eqnarray}
with the coupling coefficients
\begin{eqnarray}\nn
  \alpha&=&\frac{2}{\kappa^2},\;\;\;\;\beta=\frac{\kappa^2}{2\omega^4}, \\\nn
  \gamma&=&-\frac{\kappa^2\mu}{2\omega^2},\;\;\;\;\zeta=\frac{\kappa^2\mu^2}{8},\;\;\;\;\eta=-\frac{\kappa^2\mu^2}{32}\frac{1-4\lambda}{1-3\lambda}, \\\nn
  \xi&=&-\frac{\Lambda_W\kappa^2\mu^2}{8(1-3\lambda)},\;\;\;\;
  \sigma=\frac{3\Lambda_W^2\kappa^2\mu^2}{8(1-3\lambda)}.
\end{eqnarray}
The essential difference from the action in \cite{Horava:2009uw}
comes from $S_V^{\text{(UV)}}$.
 The dimensionless parameters $\beta_1$, $\beta_2$ and
$\beta_3$ in it are the ones dominating the UV behavior. In
detailed balance case, from \eqref{Svuv,with detailed balance}, we
have $\beta_1=1,\beta_2=-1,\beta_3=-1/8$. But here when discarding
the condition, these three parameters can be arbitrary.
Furthermore, we have made the following analytic continuation
\begin{equation}\label{continuation}
\mu\rightarrow i\mu,\hspace{5ex} \omega^2\rightarrow -i \omega^2
\end{equation}
as suggested in \cite{Lu2009} in order to have a physical
evolution. As a result, the emergent speed of light is
\begin{equation}
c=\frac{\kappa^2 \mu}{4}\sqrt{\frac{\Lambda_W}{3\lambda-1}},
\end{equation}
which requires $\Lambda_W$ being positive for $\lambda >
\frac{1}{3}$.

 The above action is not the most general one we can have.
For simplicity, we still keep most of the terms in the
\hl~gravity, but only modify the marginal terms which are
quadratic in curvature and of two spatial derivatives.

The first task is to derive the constraint equation. Write down
Hamiltonian\cite{Horava:2008ih} and take the variations with
respect to the momentum conjugate to $N_i$ and $N$, we get
respectively two equations. One of them corresponding to $N_i$
gives so called super-momentum constraint:
\begin{eqnarray}\label{Niconstraint}
  0&=&\nabla_iN(K^i{}_j-\lambda K\delta^i_j) 
\end{eqnarray}
which lead to the relation on $B$ up to the first order
\begin{equation}\label{constraint:B}
    c\p^2 B=\frac{1-3\lambda}{\lambda-1}a\dot\psi.
\end{equation}
The constraint corresponding to $N(t)$ is subtler. Since we
require projectivity condition, we cannot get  the local
super-Hamiltonian constraint. For the classical evolution, the
isotropy and homogeneity of the background allow us to reduce the
integral constraint to a local one, which is just the Friedman
equation on the scale factor:
\begin{eqnarray}\label{constraint:sigma}
  \big(\frac{\dot{a}}{a}\big)^2 =
  c^2\frac{\sigma}{3\alpha(1-3\lambda)}=\frac{c^4\Lambda_W}{3\lambda-1}.
\end{eqnarray}

From now on, our discussion will  base on the gravity action with
above analytic continued parameters. As a consequence, the Hubble
``constant" $H=\dot{a}/a$ is really a constant if we just focus on
classical evolution and the universe is undergoing a exponentially
expansion no matter whatever $\lambda$ is. In other words, the
universe is in a de-Sitter phase once $\lambda$ is fixed. This
means that the early universe is very much like in an
exponentially expanding inflation period. The key difference is
that we do not need a scalar field with slow roll potential to
drive the inflation.

Another important feature of the evolution is that the
acceleration is related to the emergent speed of light. At UV, the
speed of light could be large such that the acceleration is quite
large. But at IR, the acceleration is the same as the late time
acceleration in Einstein gravity with a positive cosmological
constant.


Using \eqref{constraint:B} and \eqref{constraint:sigma}, we
calculate the action of second order.
\begin{eqnarray}\label{Sk(2)}
{}^{(2)}S_K&=&c^{-1}\alpha\int dtd^3\mb{x}a^3
3(1-3\lambda)\left[\frac{2}{3(1-\lambda)}\dot\psi^2+6H\psi\dot\psi+\frac{9}{2}H^2\psi^2+\frac{H^2}{2}B\p^2B\right],\\\label{Svuv(2)}
{}^{(2)}S_V^{(\text{UV})}&=&-2c\beta\int
dtd^3\mb{x}\frac{1}{a^3}(3\beta_1+2\beta_2+8\beta_3)\psi\p^6\psi,\\\label{Svinter(2)}
{}^{(2)}S_V^{(\text{inter})}&=&2c(3\zeta+8\eta)\int
dtd^3\mb{x}\frac{1}{a}\psi\p^4\psi,\\\label{SvIR(2)}
{}^{(2)}S_V^{(\text{IR})}&=&c\int
dtd^3\mb{x}\left[-2a\xi\psi\p^2\psi+\frac{9}{2}a^3\sigma\psi^2-\frac{1}{2}a^3\sigma
B\p^2B\right] .
\end{eqnarray}
When adding together, the terms in ${}^{(2)}S_V^{(\text{IR})}$
involving $\sigma$ can be converted into terms of $H^2$ by virtue of
\eqref{constraint:sigma}, and can be combined with those in
${}^{(2)}S_K$. Therefore, the total action of second order is
\begin{eqnarray}\nn
  {}^{(2)}S &=& \int dtd^3\mb{x}\left\{3\frac{\alpha}{c} a^3(1-3\lambda)\left[\frac{2}{3}\frac{1}{1-\lambda}\dot\psi^2+6H\psi\dot\psi+9H^2\psi^2\right]\right. \\
   & & \left.-\frac{2\beta}{a^3}c^2(3\beta_1+2\beta_2+8\beta_3)\psi\p^6\psi+\frac{2}{a}c^2(3\zeta+8\eta)\psi\p^4\psi-2c^2a\xi\psi\p^2\psi\right\}
\end{eqnarray}


From the second order action we can derive the equation of motion,
\begin{eqnarray}\nn\label{eom:explicit}
    \ddot\psi+3H{\dot \psi}+\frac{9}{2}(1-\lambda)\dot{H}\psi
    +\frac{\kappa^4}{4\omega^4}c^2(3\beta_1+2\beta_2+8\beta_3)\frac{1-\lambda}{1-3\lambda}\frac{1}{a^6}\p^6\psi&&\\\label{eom}
    +\frac{\kappa^4\mu^4}{16}c^2\left(\frac{1-\lambda}{1-3\lambda}\right)^2\frac{1}{a^4}\p^4\psi
    +\frac{\kappa^4\mu^4}{16}c^2\Lambda_W\frac{1-\lambda}{(1-3\lambda)^2}\frac{1}{a^2}\p^2\psi&=&0.
\end{eqnarray}
It is remarkable that when $\lambda=1$, the above equation reduces
to the one in standard Einstein theory, in which case $\psi$ is
not a real dynamical field. In other words, when the theory flows
to IR fixed point where the Einstein theory and the diffeomorphism
are partially recovered, the scalar field is just a gauge artifact
due to the extra gauge degrees of freedom.

 This formula coincides with the equation of motion of
the gravitational scalar in \cite{Cai:2009dx}, except for the
$\p^6\psi$ dependent term here. We note, that the equation in
\cite{Cai:2009dx} is derived from \horava's original action with
detailed balance condition, where $\beta_1=1$,$\beta_2=-1$ and
$\beta_3=-1/8$, thus $3\beta_1+2\beta_2+8\beta_3=0$, which just
makes the $\p^6\psi$ term vanishing. Then there's no
${}^{(2)}S_V^{(\text{UV})}$ term any more, and
${}^{(2)}S_V^{(\text{inter})}$ will be dominant in UV era, when the
dispersion relation will be $E^2\sim k^4$, and spectrum will be
$\mathcal{P}_k\sim k$. On the other hand, if we discard the detailed
balance condition, the $k^6$-dependence in dispersion relation will
emerge. Moreover, one has to care about the stability and unitarity
at UV, which requires that $3\beta_1+2\beta_2+8\beta_3>0$. Finally,
following the treatment in \cite{Kiritsis:2009sh, Mukohyama:2009gg}
we can obtain the power spectrum. In UV limit, the equations of
motion reads
\begin{equation}\label{UVeom}
    \ddot{\psi}+\frac{\Omega^2}{a^6}\psi=0,\;\;\;\;
    \Omega^2=\frac{\kappa^4c^2}{4\omega^4}(3\beta_1+2\beta_2+8\beta_3)\frac{1-\lambda}
    {3\lambda-1}k^6
\end{equation}
which can be solved under WKB approximation as
\begin{equation}\label{WKBsolution}
    \psi\simeq\frac{1}{\sqrt{2\Omega}}\exp\left(-i\Omega\int\frac{dt}{a^3}\right).
\end{equation}
And we can use \eqref{WKBsolution} to calculate the scale-invariant
power spectrum,
\begin{equation}\label{spectrum}
    \mathcal{P}_\psi(k)=\frac{k^3}{2\pi^2}|\psi|^2
    =\frac{2}{\pi^2}\frac{\omega^2}{\kappa^4\mu}\frac{3\lambda-1}{\sqrt{\Lambda_W(3\beta_1+2\beta_2+8\beta_3)(1-\lambda)}}.
\end{equation}
  We see, that the power spectrum
produced by the scalar mode of the gravitational field at UV is
scale-invariant, sharing the same property
 of temperature fluctuations in CMBR observation. This is  a hint, that the gravitational scalar may be
the seed of the large scale structure of the universe and the
origin of the anisotropy of the cosmic microwave background
radiation.

It seems that the power spectrum is divergent when $\lambda \to
1$. This is an illusion. The above spectrum corresponds to the
fluctuations generated at the scale much higher than IR fixed
point. And the $\lambda$ in (\ref{spectrum}) corresponds to the
matching point between UV and IR, roughly at the horizon crossing,
at which $\lambda$ is not equal to $1$.

The fluctuations of the scalar mode would be frozen after horizon
crossing. And after IR fixed point, the scalar mode is not
physical anymore, instead it changes to the Newton potential. In
fact, to respect the projectivity condition, the perturbed FRW
metric was set to (\ref{metric}) in terms of Painleve-Gullstrand
coordinates, which is equivalent to the standard scalar perturbed
FRW metric. Moreover, the higher-derivative terms are much
suppressed after RG flows to $\lambda=1$ so that we can trust the
usual treatment in general relativity. As usual, the scalar
perturbation may combine with matter perturbation to form a
conserved curvature perturbation. Unfortunately due to the
ignorance of the coupling of matter with gravity with $z\neq 1$,
we do not know if there exist a similar quantity before RG flowing
to $\lambda=1$. It is an important issue which we would like to
address in the future.


\section{discussions}\label{sec-inclu}

In this paper we investigated the possibility that the scalar
perturbation of \hl~~ gravity, instead of matter perturbation,
seeds the large scale structure of our universe. As the first
step, we showed that the power spectrum of the gravitational
scalar at UV is scale invariant. It has been argued that in
\hl~~gravity the inflation is not necessary. Our study suggest
that we do not even need matter scalar field and gravity itself
provide the natural scalar mode for us.

We found that in order to have scale invariant spectrum, we had to
relax the detailed balance condition and have more general gravity
action. This is not a drawback. From our investigation, the
absence of Cotton tensor allows the equation of motion of scalar
mode to have spatial derivatives up to six order. Similarly this
happens for the fluctuations around the flat spacetime background.
This actually improve the ultraviolet behavior of the scalar mode.
On the other hand, the discarding of the detailed balance
condition open many other possibilities to construct the model.
Some of the possibility may have interesting physical implication.
For example, the marginal term as $R^3, R^i_jR^j_kR^k_i,
RR_{ij}R^{ij}$ may induce the cubic interaction terms of the
scalar mode, leading to non-Gaussianity\cite{CPT}.

In \hl~ gravity, to have a physical evolution requires a positive
cosmological constant from beginning. At the very early stage, the
speed of sound is very large and drive a very fast exponentially
expansion. This stage could plays the role of inflation and help
us to solve the well-known problems in big-bang cosmology. At very
late time stage, we recover the accelerating expansion due to a
small positive cosmological constant.

Even though the idea that \hl~cosmology can be an alternative to
the inflationary scenario, there are lots of questions waiting to
be answered. Due to our ignorance of the exact RG flow of the
theory, it is not clear when  and how the Einstein gravity is
recovered and the scalar mode disappear. It is also an interesting
issue to study how the gravitational scalar is related to the
temperature perturbation. More fundamentally, it is essential to
study the ultraviolet behavior of the \hl~gravity, especially
considering so many marginal terms and relevant terms without
detail balance. Besides the requirement of stability and
unitarity, does there exist other principle to determine the
parameters? It could be possible that there exist UV fixed points
characterized by a few parameters. The RG flows from UV fixed
points to IR by various relevant terms are essential for us to
understand the \hl~cosmology.

\section*{Acknowledgments}
The work was partially supported by NSFC Grant No.10535060,
10775002, NKBRPC (No. 2006CB805905) and RFDP. We would like to
thank Qing-guo Huang for stimulating questions and comments.


\begin{thebibliography}{99}
\bibitem{Quantumphase}R.H. Mornreich, M. Luban and S. Shtrikman,
``Critical Behavior at the Onset of $\vec{k}$-space Instability on
the $\lambda$ Line", Phys. Rev. Lett. {\bf 35} (1975)1678; \\
S. Sachdev, ``Quantum Phase Transitions", Cambridge U.P. (1999);\\
E. Ardonne, P. Fendley and E. Fradkin, ``Topological Order and
Conformal Quantum Critical Points", Annals Phys. {\bf 310}
(2004)493-551, [cond-mat/0311466].


\bibitem{Horava:2008ih}
  P.~Horava,
  ``Membranes at Quantum Criticality,''
  JHEP {\bf 0903}, 020 (2009)
  [arXiv:0812.4287 [hep-th]].


\bibitem{Horava:2009uw}
  P.~Horava,
  ``Quantum Gravity at a Lifshitz Point,''
  Phys.\ Rev.\  D {\bf 79}, 084008 (2009)
  [arXiv:0901.3775 [hep-th]].

\bibitem{Calcagni:2009ar}
  G.~Calcagni,
  ``Cosmology of the Lifshitz universe,''
  arXiv:0904.0829 [hep-th].

\bibitem{Kiritsis:2009sh}
  E.~Kiritsis and G.~Kofinas,
  ``Horava-Lifshitz Cosmology,''
  arXiv:0904.1334 [hep-th].

\bibitem{Mukohyama:2009gg}
  S.~Mukohyama,
  ``Scale-invariant cosmological perturbations from Horava-Lifshitz gravity
  without inflation,''
  arXiv:0904.2190 [hep-th].


\bibitem{Lifshitz}E.M. Lifshitz, ``On the Theory of Second-Order
Phase Transitions I \& II", Zh. Eksp. Teor. Fiz {\bf 11} (1941)255
\& 269.

\bibitem{Chen:2009ka}
  B.~Chen and Q.~G.~Huang,
  ``Field Theory at a Lifshitz Point,''
  arXiv:0904.4565 [hep-th].

\bibitem{Lu2009}H.~Lu, J.~Mei and C.~N.~Pope,
  ``Solutions to Horava Gravity,''
  arXiv:0904.1595 [hep-th].



\bibitem{HL}
   H.~Nastase,
  ``On IR solutions in Horava gravity theories,''
  arXiv:0904.3604 [hep-th]. R.~Brandenberger,
  ``Matter Bounce in Horava-Lifshitz Cosmology,''
  arXiv:0904.2835 [hep-th].
  H.~Nikolic,
  ``Horava-Lifshitz gravity, absolute time, and objective particles in curved
  space,''
  arXiv:0904.3412 [hep-th].
  R.~G.~Cai, L.~M.~Cao and N.~Ohta,
  ``Topological Black Holes in Horava-Lifshitz Gravity,''
  arXiv:0904.3670 [hep-th].
  R.~G.~Cai, Y.~Liu and Y.~W.~Sun,
  ``On the z=4 Horava-Lifshitz Gravity,''
  arXiv:0904.4104 [hep-th].
  Y.~S.~Piao,
  ``Primordial Perturbation in Horava-Lifshitz Cosmology,''
  arXiv:0904.4117 [hep-th].
  E.~O.~Colgain and H.~Yavartanoo,
  ``Dyonic solution of Horava-Lifshitz Gravity,''
  arXiv:0904.4357 [hep-th].
  U.~H.~Danielsson and L.~Thorlacius,
  ``Black holes in asymptotically Lifshitz spacetime,''
  JHEP {\bf 0903}, 070 (2009)
  [arXiv:0812.5088 [hep-th]].
  P.~Horava,
  ``Spectral Dimension of the Universe in Quantum Gravity at a Lifshitz
  Point,''
  arXiv:0902.3657 [hep-th].
  T.~Takahashi and J.~Soda,
  ``Chiral Primordial Gravitational Waves from a Lifshitz Point,''
  arXiv:0904.0554 [hep-th].
  X.~Gao,
  ``Cosmological Perturbations and Non-Gaussianities in Ho\v{r}ava-Lifshitz
  Gravity,''
  arXiv:0904.4187 [hep-th].
G.E. Volovik,``z=3 Lifshitz-Horava model and Fermi-point scenario
of emergent gravity", arXiv:0904.4113 [gr-qc]. T. Sotiriou, M.
Visser, S. Weinfurtner, ``Phenomenologically viable
Lorentz-violating quantum gravity",  arXiv:0904.4464 [hep-th]. S.
Mukohyama, K. Nakayama, F. Takahashi and S. Yokoyama,
``Phenomenological Aspects of Horava-Lifshitz Cosmology",
arXiv:0905.0055 [hep-th]. Y. S. Myung and Y.W. Kim,
``Thermodynamics of Ho\v{r}ava-Lifshitz black holes",
arXiv:0905.0179 [hep-th]. T. Nishioka, ``Horava-Lifshitz
Holography", arXiv:0905.0473 [hep-th]. Songbai Chen, Jiliang Jing,
``Quasinormal modes of a black hole in the deformed
H\v{o}rava-Lifshitz gravity",  arXiv:0905.1409 [gr-qc]. R.B. Mann,
``Lifshitz Topological Black Holes", arXiv:0905.1136 [hep-th].Yun
Soo Myung, ``Thermodynamics of black holes in the deformed
Ho\v{r}ava-Lifshitz gravity", arXiv:0905.0957 [hep-th].Ahmad
Ghodsi, ``Toroidal solutions in Horava Gravity",  arXiv:0905.0836
[hep-th].Rong-Gen Cai, Li-Ming Cao, Nobuyoshi Ohta,
``Thermodynamics of Black Holes in Horava-Lifshitz Gravity",
arXiv:0905.0751 [hep-th] S.Kalyana Rama, ``Anisotropic Cosmology
and (Super)Stiff Matter in Ho\v{r}ava's Gravity Theory",
arXiv:0905.0700 [hep-th].Alex Kehagias, Konstadinos Sfetsos, ``
The black hole and FRW geometries of non-relativistic gravity",
arXiv:0905.0477 [hep-th].R.A. Konoplya, ``Towards constraining of
the Horava-Lifshitz gravities", arXiv:0905.1523 [hep-th]. Robert
H. Brandenberger, ``Processing of Cosmological Perturbations in a
Cyclic Cosmology", arXiv:0905.1514 [hep-th]. D. Orlando and S.
Reffert, ``On the Renormalizability of Horava-Lifshitz-type
Gravities", arXiv:0905.0301[hep-th]. J. Kluson, ``Branes at
Quantum Criticality", arXiv:0904.1343 [hep-th]. J. Kluson,
``Stable and Unstable D-Branes at Criticality", arXiv:0905.1483
[hep-th].













\bibitem{Cai:2009dx}
  R.~G.~Cai, B.~Hu and H.~B.~Zhang,
  ``Dynamical Scalar Degree of Freedom in Horava-Lifshitz Gravity,''
  arXiv:0905.0255 [hep-th].

\bibitem{Maldacena:2002vr}
  J.~M.~Maldacena,
  ``Non-Gaussian features of primordial fluctuations in single field
  inflationary models,''
  JHEP {\bf 0305}, 013 (2003)
  [arXiv:astro-ph/0210603].
\bibitem{CPT} B, Chen, S. Pi and J.Z. Tang, Work in progress.
\end{thebibliography}
\end{document}